\documentclass[aps,pra,twocolumn,amsmath,amssymb]{revtex4-2}

\usepackage{graphicx}
\usepackage{dcolumn}
\usepackage{bm}
\usepackage{color}

\begin{document}


\title{Multiqubit Toffoli gates and optimal geometry with Rydberg atoms}
\author{Dongmin Yu$^{1}$}\thanks{First Author and Second Author contribute equally to this work.\\} 
\author{Han Wang$^{1}$}\thanks{First Author and Second Author contribute equally to this work.\\} 
\author{Jin-ming Liu$^{1}$} \author{Shi-Lei Su$^{2,\ddagger}$} \author{Jing Qian$^{1,\dagger}$} \author{Weiping Zhang$^{3,4,5}$ }

\affiliation{$^{1}$State Key Laboratory of Precision Spectroscopy, Department of Physics, School of Physics and Electronic Science, East China
Normal University, Shanghai 200062, China}
\affiliation{$^{2}$School of Physics, Zhengzhou University, Zhengzhou 450001, China}
\affiliation{$^{3}$School of Physics and Astronomy, Tsung-Dao Lee Institute, Shanghai Jiao Tong University, Shanghai 200240, China}
\affiliation{$^{4}$Collaborative Innovation Center of Extreme Optics, Shanxi University, Taiyuan, Shanxi 030006, China}
\affiliation{$^{5}$Shanghai Research Center for Quantum Sciences, Shanghai 201315, China}

\begin{abstract}
Due to its potential for implementing a scalable quantum computer, multiqubit Toffoli gate lies in the heart of quantum information processing. In this article, we demonstrate a multiqubit blockade gate with atoms arranged in a three-dimension spheroidal array. The gate performance is greatly improved by the method of optimizing control-qubit distributions on the spherical surface via evolutionary algorithm, which leads to an enhanced asymmetric Rydberg blockade. This spheroidal configuration, not only arises a well preservation for the dipole blockade energy between arbitrary control-target pairs, which keeps the asymmetric blockade error at a very low level; but also manifests an unprecedented robustness to the spatial position variations, leading to a negligible position error. Taking account of intrinsic errors and with typical experimental parameters, we numerically show that a C$_6$NOT Rydberg gate can be created with a fidelity of 0.992 which is only limited by the Rydberg state decays.
 Our protocol opens up a new platform of higher-dimensional atomic arrays for achieving multiqubit neutral-atom quantum computation.
\end{abstract}

\email{jqian1982@gmail.com}
\maketitle
\preprint{}

\section{Introduction}
Rydberg atoms serve as a reliable platform for studying quantum computing and quantum simulation because of their strong and tunable interactions, which can block the excitation of surrounding atoms in the vicinity of a preexcited atom \cite{Urban2009,Alpha2009,Ravets2014}. Via this so-called Rydberg blockade mechanism, versatile quantum gates can be created \cite{PhysRevLett.104.010503,PhysRevLett.85.2208,PhysRevLett.87.037901,PhysRevA.82.030306,PhysRevA.89.032334} which manifest as basic logic-calculation units for universal quantum computation \cite{RevModPhys.82.2313,nielsen_2010}. Among existing Rydberg-mediated quantum gates, a multiqubit Toffoli (C$_n$NOT) gate is an important family member, which can offer an efficient implementation of Grover quantum search algorithm to speedup the searches on a programmable quantum computer \cite{Figgatt2017} or to extend into any dimensional quantum systems \cite{saha2021asymptotically}. A conventional three-qubit Toffoli (C$_2$NOT) gate can be implemented in a 1D array \cite{PhysRevA.76.022334,PhysRevA.98.042704} where two outer control atoms constrain the behavior of middle target atom with strong control-target interactions. However such a linear model
is unsuited for engineering a multiqubit gate because one target atom can not be simultaneously manipulated by two nearest-neighbor control qubits due to the blockade mechanism \cite{PhysRevA.77.032723}. 
Therefore previous contributions to a multiqubit Toffoli gate often rely on the assembly of several elementary gates \cite{PhysRevA.52.3457,PhysRevA.87.062318,Ralph_2007,Yu_2013,Biswal_2019} or the parallel operation on some clusters of atoms in a 1D array of optical tweezers \cite{Levine_2019}. Direct execution of multiqubit Toffoli gates ($n \geq3$) remains a big challenge both in theory and experiment.

To date, several studies for multiqubit gates use the way of adiabatic passages in which the evolution of states can be performed by obeying a multiqubit dark eigenstate with complex optimal pulses \cite{PhysRevA.103.062607,PhysRevX.10.021054}. An alternative way for this target depends on nonadiabatic holonomic quantum computation showing a C$_3$NOT gate with an error of 0.0018 \cite{PhysRevA.104.012618}.
Another prominent idea to the realization of multiqubit Rydberg gates adopts asymmetric blockade as proposed in \cite{Saffman_2009}, in which there exists a large separation of scales between different types of Rydberg interactions \cite{Isenhower_2011,Wu_2010,Su_2018,PhysRevLett.127.120501}. However we note that, the asymmetric interaction condition breaks easily when the number of qubits is enlarged, especially for atoms arranged in 1D or 2D arrays where distant control-target interaction suffers from a dramatic decrease. Recently J. Young and coworkers propose a 2D multiqubit gate by placing many control and many target atoms at the same time, 
in which the strong control-control and control-target interactions can be engineered via extra microwave fields, leading to 
a perfect asymmetric blockade \cite{PhysRevLett.127.120501}. But this scheme is still unsuitable for implementing individual multicontrol \cite{PhysRevA.98.032306,PhysRevA.75.034307,PhysRevA.103.052437} or multitarget \cite{PhysRevA.101.022308} gates due to the absence of strong and tunable interactions between distant atoms.

 In the present work, inspired by the development of defect-free atom arrays from 2D to 3D platforms where arbitrary atoms can be arranged expectantly in space \cite{Barredo2018,Kumar2018,PhysRevLett.115.043003,Barredoaah3778,PhysRevLett.119.180503,PhysRevLett.122.203601} (a recent work has reported mixed-species atom arrays with arbitrary geometry \cite{PhysRevLett.128.083202}), we propose a scheme for implementing C$_n$NOT gates with atoms individually arranged in a 3D spheroidal atomic array. As illustrated in Fig.\ref{mod}, we consider a single target atom(in green) located at the center and $n$ control atoms(in red) on the surface. Such a 3D atomic array can be treated as an assembling of multi-layer 2D lattices and easily achieve single-site Rydberg addressing \cite{JPB53}. 
 Compared to the existing asymmetric-blockade-based protocols \cite{Saffman_2009,Su_2018,PhysRevLett.127.120501,Wu_2010,Isenhower_2011}, our scheme benefits from an optimal 3D configuration to maximize the asymmetry of blockade, representing an unprecedented robustness to the 3D atomic position variations. 
 The synthetic interplay between interatomic Rydberg-Rydberg interactions and the optimal geometry results in a huge asymmetric blockade, making the gate imperfection dominated by a intrinsic decay error, and the asymmetric blockade error can be suppressed to a negligible level. Our results show that a simple estimate of decay errors gives rise to an acceptable fidelity of 0.9537 for a multiqubit C$_{12}$NOT gate.
 This 3D Rydberg quantum gates can serve as a new gate-unit for parallel operation in 3D optical tweezers, promising for scalable quantum computation with more flexibility.


 \section{Maximizing asymmetric blockade via optimization}
 \label{MAX}

 \begin{figure}
\centering\includegraphics[width=0.48\textwidth]{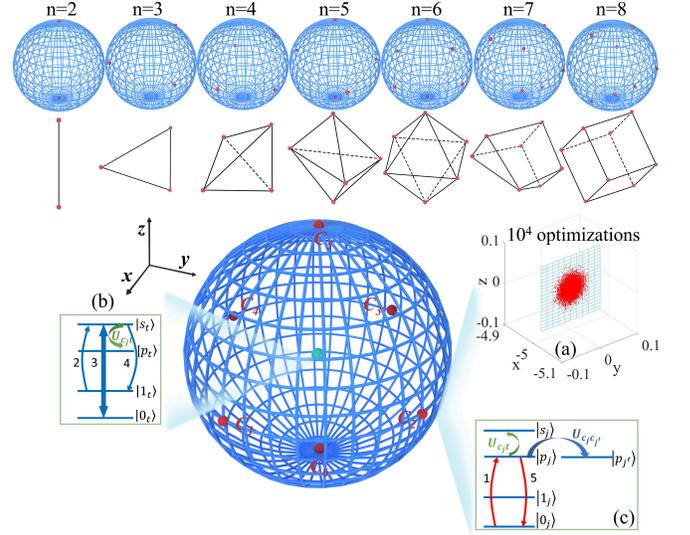}
\caption{Realization of spheroidal multiqubit Toffoli gates in a 3D atomic array. Upper panels: optimized distributions of $n$ control atoms on the surface accompanied by the optimal geometries explicitly shown below. Main panel: Amplification of the C$_6$NOT gate with four control atoms $c_{1\sim4}$ at the equatorial plane and two control atoms $c_{5\sim6}$ at the south and north poles. The target atom is placed at the center. (a) Distribution of $c_2$ after $10^4$ optimizations via evolutionary algorithm. (b-c) The atomic energy levels as well as the atom-light interactions. In the presence of a static electric field, we fix the quantization axis along $+\hat{z}$ to simplify the optimization, which arises an angular-dependent dipole-dipole interaction $U_{c_jt}(\theta_{c_jt})$ for each control-target atomic pair, while the control-control interaction $U_{c_jc_j^{\prime}}(|\bold{r}_{j}-\bold{r}_{j^{\prime}}|)$ is of {\it vdWs}-type which depends on the intraspecies distance $R_{c_jc_j^\prime}$.}
\label{mod}
\end{figure}

 To achieve desirable asymmetric interactions we adopt optimization with evolutionary algorithm. Since each atom contains two Rydberg states $|p_{j,t}\rangle$ and $|s_{j,t}\rangle$ with subscript $j(t)$ for the control(target) atom, we consider the interaction between atoms in product states $|p_js_t\rangle$ and $|s_jp_t\rangle$ is of resonant dipole-dipole feature \cite{D_yachkov_2016}
 \begin{equation}
   \hat{U}_{c_jt}(\theta_{c_jt}) =
\frac{C_3^{sp}(\theta_{c_jt})}{R_{ct}^3}(|p_js_t\rangle\langle s_jp_t|+|s_jp_t\rangle\langle p_js_t|),
\label{uct}
 \end{equation}
 with $C_3^{sp}(\theta_{c_jt})=C_3(1-3\cos^2\theta_{c_jt})$ and $C_3=|\mu_{sp}|^2/(8\pi\epsilon_0)$ (see Appendix A for more details). Note that the spheroidal structure can preserve the control-target distance $R_{ct}$ unchanged so  $U_{c_jt}$ only depends on the polarizing angle $\theta_{c_jt}$. 
 On the other hand there is also a {\it vdWs} interaction between two control atoms, given by \cite{PhysRevLett.110.263201}
 \begin{equation}
     \hat{U}_{c_jc_{j^{\prime}}}(|\bold{r}_j-\bold{r}_{j^\prime}|) =
 \frac{C_6}{R_{c_jc_{j^{\prime}}}^6}|p_jp_{j^{\prime}}\rangle\langle p_jp_{j^{\prime}}|,
 \label{ucc}
 \end{equation}
 with $R_{c_jc_{j^{\prime}}}=|\bold{r}_j-\bold{r}_{j^\prime}|$ the control-control distance, which is a {\it vdWs} energy shift of the pair state $|p_jp_{j^\prime}\rangle$ considered arising from a second-order approximation of the nonresonant dipole-dipole interaction.

 Below we focus on how to achieve best asymmetric blockade using $^{87}$Rb Rydberg states: $|s_{t,j}\rangle=|(m+1)S_{1/2},m_j=\frac{1}{2}\rangle$, $|p_{t,j}\rangle=|mP_{3/2},m_j=\frac{3}{2}\rangle$ as in \cite{Isenhower_2011}.
 The $C_3$ coefficient for $m=60$ is about $C_3/2\pi=4.194$ GHz$\cdot\mu$m$^3$
 and $C_6/2\pi=-12.0$ GHz$\cdot\mu$m$^6$ calculated by the ARC open source library \cite{SIBALIC2017319}.
 Since $U_{c_jt}\propto (1-3\cos^2\theta_{c_jt})$ and $U_{c_jc_{j^{\prime}}}\propto1/R_{c_jc_{j^{\prime}}}^6$, the condition of strongly asymmetric interactions $U_{c_jt}\gg U_{c_jc_{j^{\prime}}}$ {\it i.e.} any control atom can block the excitation of the target atom without blocking other control atoms, can be readily met if $R_{c_jc_j^{\prime}}$ is appropriate. \textcolor{black}{In Appendix B we verify the establishment of strong asymmetric interactions by calculating the leakage error due to nonresonant Rydberg couplings nearby}.
 We also note that the asymmetry increases for small principal quantum number because the coefficient $C_3$($C_6$) scales as $\sim m^4$($\sim m^{11}$) \cite{Saffman_2009}. Lowering $m$ can realize a 3D quantum gate with more control qubits (details in Sec. V).

Here, in order to maximize asymmetric blockade, we have to optimize the spatial positions of all control atoms accompanied by
a dipole-angle optimization. For arbitrary control atom $c_j$ a factor characterizing asymmetry is defined as
\begin{equation}
    \chi_j=U_{c_jt}/U_{c_jc_{j^{\prime}}},
\end{equation}
 which must be maximized. Intuitively, as increasing $n$ the {\it vdWs} interaction is enhanced so as to easily break the asymmetry. To determine maximal $n_{max}$ permitted for a chosen radius $R_{ct}$(the scale of 3D array), we perform a global optimization to the atomic positions via 
 evolutionary algorithm \cite{Cheng2018}. A detailed description of optimization algorithm can be found in Appendix C. For achieving strong asymmetric blockade, we set $\chi_j > 100$ which means the minimal value of $\chi_j$ should satisfy 
 $\min(\chi_j)>100$ for any $c_j$. This limitation leads to $n_{max}=8$ when $R_{ct}=5.0$ $\mu$m and $m=60$.
 Several optimal geometries are shown in Fig. \ref{mod}(upper panels) where the positions of control atoms denoted as red dots, are precisely obtained by sufficient optimization. This optimal structure does not depend on the coefficients $C_3$ or $C_6$ chosen and is stably existing. For an even $n$ value, the structure looks more regular. Physics behind these optimal geometries can be understood by seeking for a maximal asymmetry between dipole-dipole interaction and {\it vdWs} interaction, where the potential energy of system reaches its global minimum, corresponding to a maximal magnitude of dipole-dipole interaction. This specific geometry is formed by a competition between attractive {\it vdWs} interactions and inhomogeneous dipole-dipole interactions which is discussed in Appendix D.  Fig.\ref{mod}(a) represents an amplified position distribution of $c_2$ under $10^4$ optimization. They are extremely condensed in space, confirming the accuracy of algorithm.



 \section{Gate performance and decay error}
 As examples we investigate the gate performance of an optimal 3D C$_6$NOT gate. The effective non-Hermitian Hamiltonian
including the dissipative dynamics, is expressed as
\begin{equation}
\hat{\mathcal{H}}_{ eff}=\hat{\mathcal{H}}_0+\hat{\mathcal{H}}_I-\frac{i}{2}\sum_k\hat{\mathcal{L}}_k^{\dagger}\hat{\mathcal{L}}_k,
\end{equation}
with $k$ the indices of Rydberg levels $|p_{j,j^{\prime},t}\rangle$ and $|s_{j,t}\rangle$, and the Hamiltonians 
\begin{eqnarray}
\hat{\mathcal{H}}_0&=& \frac{1}{2}\{\Omega_{c}\sum_{j}^n(
|0_j\rangle\langle p_j|+|p_j\rangle\langle 0_j|) \label{single}\\
&+&\Omega_{t}[|1_t\rangle\langle s_t|+|s_t\rangle\langle 1_t|+(|0_t\rangle\langle s_t|+H.c.)]\}, \nonumber\\
\hat{\mathcal{H}}_I&=&\sum_{j,j^{\prime}>j}
U_{c_jc_{j^{\prime}}} + \sum_j U_{c_jt} \label{inter}
\end{eqnarray}
represent the atom-light couplings and the atom-atom Rydberg interactions. $\Omega_{c(t)}$ is the Rabi frequency for the control(target) atoms. To characterize the gate performance
we calculate the average gate fidelity
\begin{equation}
\begin{aligned}
\bar{\mathcal{F}}_n=\frac{1}{2^{n+1}}\text{Tr}\{[\sqrt{\rho_{et}}|\bar{\Psi}_{out}\rangle\langle\bar{\Psi}_{out}|\sqrt{\rho_{et}}]^{1/2}\}
\label{fide}
\end{aligned}
\end{equation}
by solving the stochastic Schr\"{o}dinger equation subject to arbitrary computational basis $|\Psi\rangle$ \cite{Molmer_1993}:
\begin{equation}
    \partial_t|\Psi\rangle=-i\hat{\mathcal{H}}_{eff}|\Psi\rangle.
    \label{she}
\end{equation}
 During each time interval $\delta t$, one generates a random number $\delta p$ and compares it with the instantaneous population on Rydberg states. If $\delta p$ is larger, the system will evolve by obeying the Schr\"{o}dinger equation (\ref{she}); otherwise one generates a random Rydberg excitation via a quantum jump \cite{PhysRevLett.108.023602}. The total random number is $t_{det}/\delta t$ where
$t_{det}=2\pi/\Omega_c+3\pi/\Omega_t$ is the gate duration. By initializing $2^{n+1}$ input states, $\bar{\Psi}_{out}$ denotes the average output at $t=t_{det}$ after 500 stochastic evolutions and $\rho_{et}$ is an etalon matrix.
In addition, the operator
\begin{equation}
 \hat{\mathcal{L}}_{k}=\sqrt{\Gamma_k}(|1\rangle\langle k|+|0\rangle\langle k|) 
\end{equation}
indicates the spontaneous population decay of Rydberg levels, in which the decay rates are $\Gamma_k=\Gamma_p$ for $k=p_{j,j^{\prime},t}$ and $\Gamma_k=\Gamma_s$ for $k=s_{j,t}$. 


\begin{figure}
\centering
\includegraphics[width=0.48\textwidth]{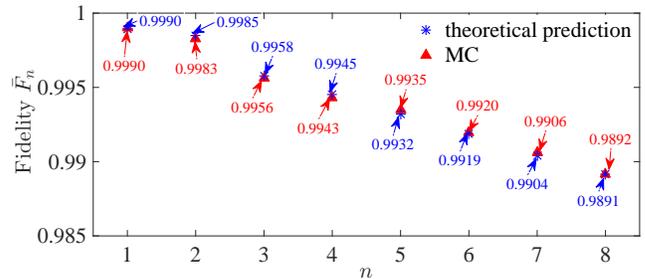}
\caption{Average gate fidelity $\bar{\mathcal{F}}_n$ as a function of the control atomic number $n$, estimated by the numerical Monte Carlo method(red triangles) and the theoretical expression(blue stars).}
\label{fed}
\end{figure}

By performing further calculations, for $n=6$ we find $(\min(U_{c_{jt}}),\max(U_{c_jc_j^{\prime}}))/2\pi = (33.552,0.096) $ MHz, leading to the asymmetry: $\min(\chi_j) = 349.5>100$. Such a huge asymmetry can keep the intrinsic asymmetric error originating from imperfect control-target(control) (anti)blockade at a very low level $< 10^{-5}$. In turn we extend this asymmetric condition to a more generalized form, as
\begin{equation}
 \min(U_{c_jt}),\Omega_c\gg\Omega_t\gg \max(U_{c_jc_{j^{\prime}}}),
 \label{ass}
\end{equation}
which is also related to relevant pulse strengths \cite{Saffman_2009}. Based on Eq.(\ref{ass}) we assume
\begin{equation}
  \Omega_t=\min(U_{c_jt})/20,\Omega_c=5\min(U_{c_jt})  
  \label{ome}
\end{equation}
throughout the paper. The decay rate is $\Gamma_{s}=5.0$ kHz and $\Gamma_{p}=3.4$ kHz in a cryogenic environment \cite{PhysRevA.102.042607}, we find a gate fidelity of $\bar{\mathcal{F}}_6=0.9920$ which is mainly constrained by the decay error from Rydberg levels(the decay error is about $8\times 10^{-3}$ estimated by Eq.(\ref{seq})). 
The overall gate time is $t_{det}\approx3\pi/\Omega_t=894$ ns. 
Detailed description of the gate operation can be found in Appendix E.
In Fig.\ref{fed} we show that the average gate fidelity $\bar{\mathcal{F}}_n$(red triangles and texts, estimated by MC) decreases with the control atom number $n$. For comparison, it is instructive to recall the decay-error expression \cite{Isenhower_2011}
\begin{equation}
    \mathcal{E}_{n,se}=\frac{1}{2^n}\frac{\pi\Gamma_s}{\Omega_t}+n\frac{3\pi\Gamma_p}{2\Omega_t}+n\frac{\pi\Gamma_p}{2\Omega_c},
    \label{seq}
\end{equation}
by which the gate fidelity can be analytically obtained according to $\bar{\mathcal{F}}_n\approx1-\mathcal{E}_{n,se}$(blue stars and texts). A good agreement is observable between the theoretical and numerical predictions which confirms that other intrinsic asymmetric errors including blockade error and antiblockade error, are both negligible due to the huge asymmetry in our scheme.

\section{Resilience to position variations}
Owing to the finite temperature which leads to atomic position variations in the optical trap, the interatomic interaction strength is slightly different for each measurement. 
This so-called {\it position error} could catastrophically break the implementation of Rydberg antiblockade(RAB)-based gates which depend on a severely modified RAB condition
\cite{Shao:14,Wu:21,PhysRevA.95.022319,surab20}. Although the excitation annihilation as reviewed in \cite{Browaeys2020} or transition slow-down effect \cite{PhysRevApplied.14.054058} makes blockade gates benefited from a robustness against interaction fluctuations, most current achievements are still constrained to fewer-qubit gates \cite{PhysRevLett.119.173402,PhysRevApplied.17.024014} because the blockade strength decreases significantly for two distant atoms. Here we express the control(target) atom position as
\begin{equation}
  \bold{r}_{j(t)}= \bold{r}_{0,j(t)}+\delta\bold{r}_{j(t)},
\end{equation}
where $\bold{r}_{0,j}=(R_{ct},\theta_{c_jt},\phi_j)$ is obtained by optimization and  $\bold{r}_{0,t}=(0,0,0)$. The displacements $\delta\bold{r}_{j(t)}$ originating from thermal motion of atoms, can be modeled as a 3D Gaussian function with widths
$\sigma_{x,y,(z)}=\sqrt{k_BT_a/mw_{x,y,(z)}^2}$ for radial(axial) localizations. Inspired by the experimental data in Ref. \cite{Graham_2019} we consider two cases: $\sigma_x\in[0,2.0]$ $\mu$m, $\sigma_{y,z}=0.27$ $\mu$m and $\sigma_z\in[0,2.0]$ $\mu$m, $\sigma_{x,y}=0.27$ $\mu$m. For Rb atoms held at a low temperature $T_a=10$ $\mu$K, the optical trap with frequencies $2\pi\times18.22$ kHz and $2\pi\times2.46$ kHz arise a position uncertainty of 0.27 $\mu$m and $2.0$ $\mu$m, respectively.
To estimate the errors from 3D position variations we also use the way of stochastic Schr\"{o}dinger equation and obtain the numerical solution by averaging over 500 independent trajectories.

\begin{figure}
\centering
\includegraphics[width=0.49\textwidth]{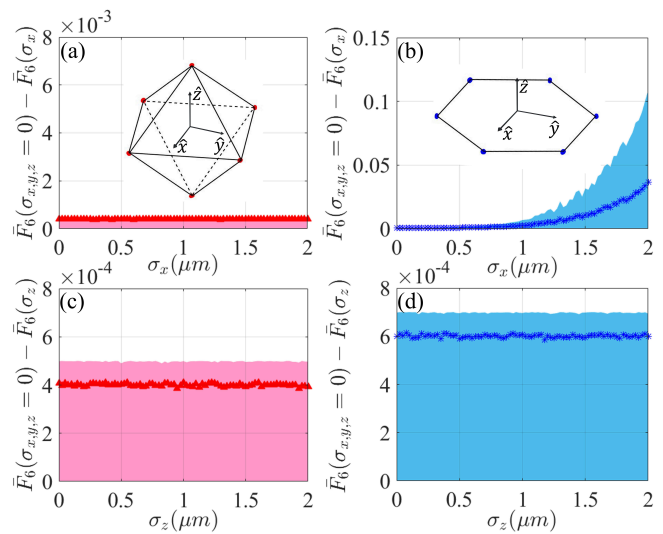}
\caption{Imperfection of the gate fidelity based on two different geometries
{\it vs} the position variations along (a-b) $\hat{x}$ and (c-d) $\hat{z}$ directions. The standard deviations are $\sigma_{y,z}=0.27$ $\mu$m in (a-b) and 
$\sigma_{x,y}=0.27$ $\mu$m in (c-d). Each point denotes an average of 500 measurements.
(a,c)(or (b,d)) are obtained from an optimal 3D C$_6$NOT gate(a 2D honeycomb-type (6+1) CNOT gate). The shadings indicate a maximal position error during the calculation.}
\label{sx}
\end{figure}

The numerical solutions in Fig.\ref{sx}(a) and (c) indicate that the 3D gate protocol can show an unprecedented robustness to the fluctuated interactions in all directions. Because in a 3D optimal configuration the position variations of atoms can be partially overcome keeping the infidelity at a small level of $10^{-4}$. In contrast, arranging $(6+1)$ atoms in a 2D honeycomb lattice will lead to a clear enhancement of the infidelity as shown in Fig.\ref{sx}(b) and (d). Especially for the radial fluctuation the imperfection dramatically increases with $\sigma_x$, agreeing with previous results \cite{Graham_2019,PhysRevLett.118.063606,PhysRevA.104.012615}. Note that a 1D chain model can not preserve the asymmetric blockade condition so as to be unable to engineer a multiqubit quantum gate. Other technical imperfections such as the sensitivity to motional dephasing, laser intensity noise and laser phase noise would be discussed in Appendix F. \textcolor{black}{A specific discussion for the leakage error due to off-resonantly coupled Rydberg pair states, will be given in Appendix B.}

\section{Large-scale multiqubit gate}

\begin{figure}
\centering
\includegraphics[width=0.485\textwidth]{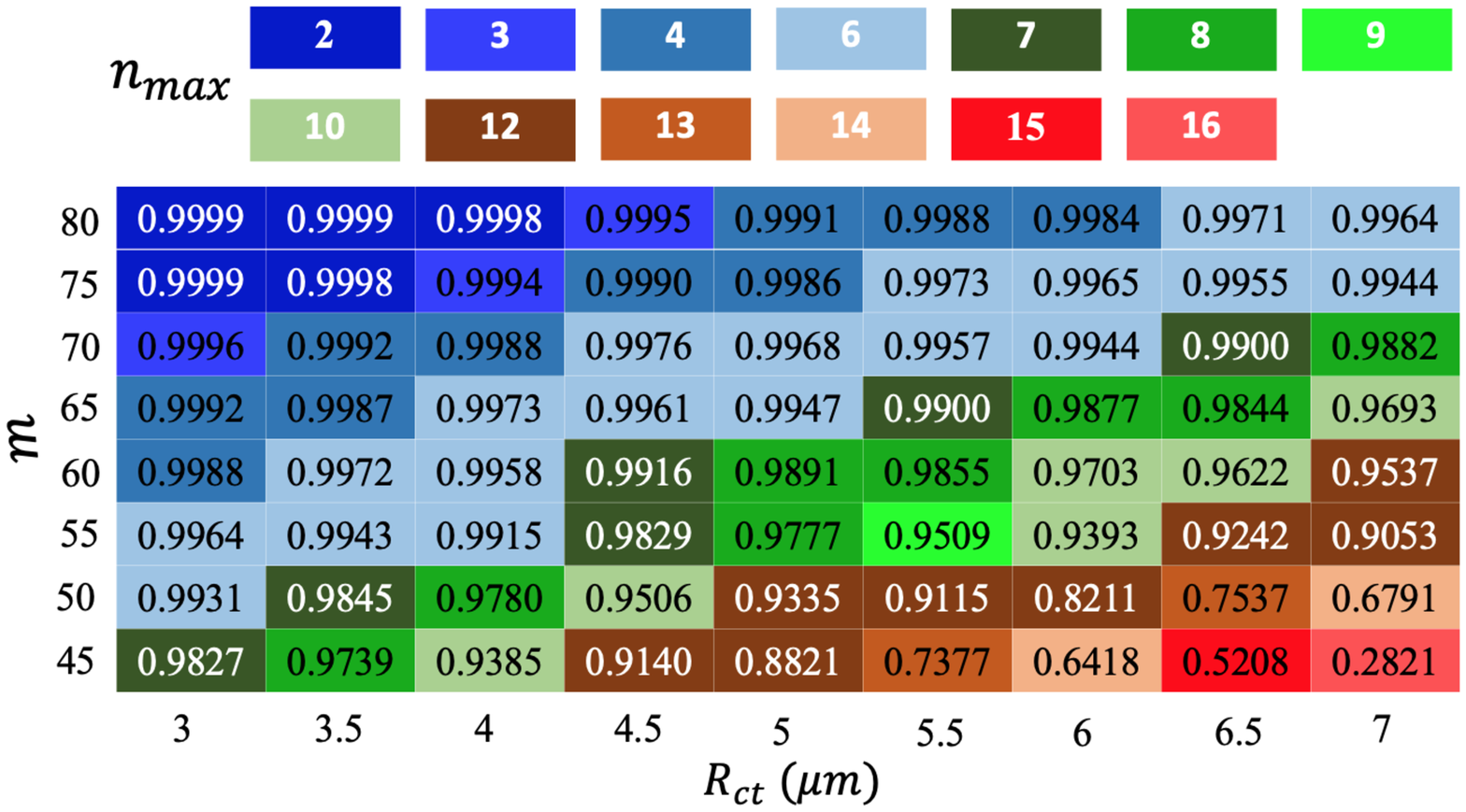}
\caption{Estimates of maximal control-atom number $n_{max}$(top, color blocks) together with the average gate fidelity $\bar{\mathcal{F}}_n$(black texts, estimated by $1-\mathcal{E}_{n,se}$) in the space of $(R_{ct},m)$. For different principal quantum number $m=(45,50,55,60,65,70,75,80)$ and at a temperature of 77 K,
the calculated coefficients are $C_6/2\pi=(-1,-2,-6,-12,-24,-57,-137,-288)$ GHz$\cdot\mu$m$^6$, $C_3/2\pi=(1.370,1.950,2.912,4.194,5.859,7.976,10.620,13.873)$ GHz$\cdot\mu$m$^3$, the Rydberg decay rates are $\Gamma_{s}=(11.70,8.55,6.53,5.00,4.09,3.33,2.76,2.33)$ kHz and $\Gamma_{p}=(6.85,5.50,4.26,3.40,2.73,2.25,1.88,1.60)$ kHz. All parameters are taken from the ARC open source library \cite{SIBALIC2017319}.}
\label{scab}
\end{figure}

This multiqubit Toffoli gates can be treated as a new calculation unit for large-scale quantum information processor \cite{Ebadi2021}. Compared to traditional fewer-qubit gates \cite{PhysRevA.85.042310,PhysRevLett.112.040501,PhysRevA.96.042306,PhysRevA.92.022336,PhysRevApplied.7.064017} our protocol benefits from an optimal 3D geometry with arbitrary $n$($<n_{max}$) control atoms. To fully determine maximal number $n_{max}$ we graphically study it by tuning the principal quantum number $m$ and the spherical radius $R_{ct}$ simultaneously. A shown in Fig.\ref{scab} it is clear that $n_{max}$ increases by lowering $m$ since the asymmetry of interactions increases then, yet at the expense of the gate fidelity. Because the decay rates $\Gamma_{s(p)}$ grow at the same time, leading to a larger decay error. On the other hand, we find $n_{max}$ has a dramatic increase with the radius $R_{ct}$. Because for a same $m$ the absolute values $\Omega_t$ and $\Omega_c$ which depend on the control-target interaction $U_{c_jt}$[Eq.~(\ref{ome})], would strongly decrease if $R_{ct}$ is enhanced. A smaller laser Rabi frequency will elongate the gate operation time, making the decay error dominant. So for $R_{ct}=7$ $\mu$m a C$_{16}$NOT gate suffers from a very low fidelity of 0.2821 when $m=45$. A high-fidelity multicontrol Rydberg-blockade gate can be accomplished by a dual consideration of both the asymmetry and the Rydberg-state decay. Our theoretical estimation shows that a C$_{12}$NOT(12 control qubits and 1 target qubit) gate with a fidelity of 0.9537 is possible when $R_{ct}=7$ $\mu$m and $m=60$.

\section{Conclusion and outlook}

We have studied a protocol of multicontrol-qubit Toffoli gates in which all control atoms are precisely arranged on a 3D spherical surface via optimization, which ensures a best asymmetric Rydberg blockade. These optimal geometries(see Fig. \ref{mod}) are obtained by performing sufficient optimization based on evolutionary algorithm. Such a spheroidal gate has many advantages.
First, it allows for a perfect preservation of strong Rydberg blockade between any control-target atom pairs, avoiding the effect of dramatic reduction in blockade strength due to distant control-target atoms as in 1D or 2D arrays. Second, 
an efficient optimization can ensure best asymmetric Rydberg blockade leading to a negligible asymmetric blockade error. Finally and most importantly, an unprecedented insensitivity $\sim10^{-4}$ to the position variations can be observed within the 3D gate due to the compensation of three-dimensional spacial fluctuations. \textcolor{black}{In comparison, Ref.\cite{Graham_2019} reports a position error of $2.5\times10^{-3}$ for a radius deviation of 0.16 $\mu$m via excitation annihilation mechanism.}
Our work shows a minimal error of 0.0463 when 12 control atoms monitor one target atom.

The scheme for an arbitrary $(n+1)$-qubit Toffoli gate can offer a direct route to multiqubit quantum computation. Upon the basis of one-step implementation to fewer-qubit quantum gates by our group recently \cite{PhysRevApplied.17.024014}, this 3D blockade-gate scheme can be used to reduce the number of fewer-qubit gates, greatly lowering the complexity of quantum device design \cite{Yu_2013,Biswal_2019}.
Other straightforward applications with multiqubit gates refer to the production of Rydberg-mediated entanglement between two atom qubits \cite{Graham_2019} or within a mesoscopic ensemble of atoms \cite{PhysRevLett.102.170502}, and of fast quantum computation with neutral Rydberg qubits \cite{PhysRevApplied.11.044035} .

\acknowledgements

This work is supported by the National Key Research and Development Program of China under Grant No. 2016YFA0302001; by the National Natural Science Foundation of China under Grants Nos. 12174106, 11474094, 11104076, 11804308, 91950112, 11174081; the Science and Technology Commission of Shanghai Municipality under Grant No. 18ZR1412800,  and the ECNU Academic Innovation Promotion Program for Excellent Doctoral Students under Grant No. YBNLTS2019-023.

\appendix

\section*{Appendix A: Asymmetric interactions}
\label{asA}

 Taking account of the scheme feasibility we present details on the Rydberg pair states and their interactions in order to show the establishment of asymmetric interactions. 
We assume the two Rydberg states of each atom which are $|s_{t,j}\rangle=|61S_{1/2},\frac{1}{2}\rangle$, $|p_{t,j}\rangle=|60P_{3/2},\frac{3}{2}\rangle$. In the presence of a static electric field, when two atoms(control and target) are prepared in two different dipole-coupled Rydberg states such as $|p_j\rangle$ and $|s_t\rangle$, the pair state $|p_js_t\rangle$ is directly coupled to the same-energy state $|s_jp_t\rangle$ by a resonant dipole-dipole exchange interaction. Typically this dipole-dipole interaction $\mathcal{B}$ between a pair of Rydberg atoms can be given by \cite{PhysRevLett.93.153001} \begin{equation}
    \mathcal{B}=\frac{1}{4\pi\varepsilon_0}[\frac{\boldsymbol{\mu}_1\cdot\boldsymbol{\mu}_2}{R^3}-3\frac{(\boldsymbol {\mu}_1\cdot\boldsymbol{R})(\boldsymbol {\mu}_2\cdot\boldsymbol{R})}{R^5}],
\end{equation}
where $\boldsymbol{\mu}_{1,2}$ stands for the electric dipole transition operators and $\varepsilon_0$ the permittivity of a vacuum. $\boldsymbol{R}$ is the internuclear distance and $R=|\boldsymbol{R}|$.
Moreover an external static electric field $\Vec{E}$ defines the quantization axis $\hat{z}$ which controls the orientation of the dipole moments relative to the separation vector $\boldsymbol{R}$, yielding an anisotropic dipole-dipole interaction,
\begin{equation}
    \begin{split}
    &\mathcal{B}(\theta_{c_jt})=\\
    &\frac{1}{4\pi\varepsilon_0R^3}\{\mu_{1+}\mu_{2-}+\mu_{1-}\mu_{2+}+\mu_{1z}\mu_{2z}(1-3\cos^2\theta_{c_jt})\\
    &-\frac{3\sin^2\theta_{c_jt}}{2}(\mu_{1+}\mu_{2+}+\mu_{1+}\mu_{2-}+\mu_{1-}\mu_{2+}+\mu_{1-}\mu_{2-})\\
    &-\frac{3\sin\theta_{c_jt} \cos\theta_{c_jt}}{\sqrt{2}}(\mu_{1+}\mu_{2z}+\mu_{1-}\mu_{2z}+\mu_{1z}\mu_{2+}+\mu_{1z}\mu_{2-})\}, 
    \label{spp}
    \end{split}
\end{equation}
with $\theta_{c_jt}$ the polarizing angle between the internuclear axis and the quantization axis $\hat{z}$. $\mu_{v,(x,y,z)}$ denotes the projections of dipole matrix element $\boldsymbol{\mu}_v$ onto axis $\hat{x},\hat{y},\hat{z}$ and $\mu_{v,\pm}=\mu_{v,x}\pm i\mu_{v,y}$ with $v\in(1,2)$. Accounting for the use of $\sigma$-polarized transition between $|60P_{3/2},m_j=3/2\rangle$ and $|61S_{1/2},m_j=1/2\rangle$ with respect to $\Delta m_j=\pm1$, we can ignore the term $\mu_{1z}\mu_{2z}$ which requires $\Delta m_j=0$. Then Eq.(\ref{spp}) can be reorganized as
\begin{eqnarray}
    &\mathcal{B}&(\theta_{c_jt}) \nonumber\\
    =&-&\frac{(1-3\cos^2\theta_{c_jt})}{8\pi\varepsilon_0R^3}(\mu_{1+}\mu_{2-}+\mu_{1-}\mu_{2+}) \nonumber \\
    &-&\frac{3\sin^2\theta_{c_jt}}{8\pi\varepsilon_0R^3}[\mu_{1+}\mu_{2+}+\mu_{1-}\mu_{2-}    \label{reg} \\
    &+&\frac{\sqrt{2}\cos\theta_{c_jt}}{\sin\theta_{c_jt}}(\mu_{1+}\mu_{2z}+\mu_{1-}\mu_{2z}+\mu_{1z}\mu_{2+}+\mu_{1z}\mu_{2-})]. \nonumber
\end{eqnarray}

Apparently, there are three types of angular dependence in Eq.(\ref{reg}) while only the first term $\propto(1-3\cos^2\theta_{c_jt})$ is appropriate. This corresponds to a resonant exchange energy between states $|s_{j}p_{t}\rangle\leftrightarrows|p_{j}s_{t}\rangle$ where $\Delta m_j=+1$ for one atom and $\Delta m_j=-1$ for the other. Other possible transitions connecting with
same combinations of $\Delta m_j=\pm 1$, are off-resonantly coupled due to a big Stark shift via the electric field \cite{PhysRevLett.93.153001}. \textcolor{black}{An estimation of the leakage error to the gate fidelity from these nonresonant Rydberg levels is illustrated in Appendix B. }

In the main text we expert a resonant dipole-dipole interaction strength that only varies as
\begin{equation}
 \langle s_{j(t)}p_{t(j)}|\mathcal{B}(\theta_{c_jt})|p_{j(t)}s_{t(j)}\rangle=\frac{C_3^{sp}(\theta_{c_jt})}{R_{ct}^3},
\end{equation}
where $R_{ct}=R$ means the two-atom separation and the interaction coefficient $C_3^{sp}$ scaling as $m^4$ takes a complex form of
\begin{equation}
    \begin{split}
    C_3^{sp}(\theta_{c_jt}) &=\frac{|\mu_{sp}|^2(1-3\cos^2\theta_{c_jt}) }{8\pi\varepsilon_0}, 
    \end{split}
\end{equation}
and the transition matrix element is
\begin{eqnarray}
    \mu_{sp}=\langle 61S_{1/2},m_j=1/2|\mu|60P_{3/2},m_j=3/2\rangle. \nonumber
\end{eqnarray}

Finally, we can obtain the electric dipole-dipole Hamiltonian between a pair of control and target atoms, which is
\begin{equation}
   \hat{U}_{c_jt}^{sp}(\theta_{c_jt}) =
\frac{C_3^{sp}(\theta_{c_jt})}{R_{ct}^3}(|p_js_t\rangle\langle s_jp_t|+|s_jp_t\rangle\langle p_js_t|).
\label{Uctsp}
\end{equation}
 
On the other hand, as for two control atoms which are prepared in same Rydberg level such as $|60P_{3/2},m_j=3/2\rangle$ the electric dipole-dipole interaction $\mathcal{B}$ only plays roles at the second-order in perturbation theory since an atomic state has a vanishing average electric dipole moment to the first-order of perturbation \cite{PhysRevA.75.032712}. As a result via $\mathcal{B}$ the pair state $|p_jp_{j^\prime}\rangle=|60P_{3/2},60P_{3/2}\rangle$ is coupled to other nearby pair states of opposite parity where the energy of those states differs from that of $|60P_{3/2},60P_{3/2}\rangle$ by a big quantity. 
The average effect gives rise to a second-order {\it vdWs} shift of the considered pair state $|p_jp_{j^\prime}\rangle$ scaling as $\propto C_6/R_{c_jc_j^\prime}^6$ where the coefficient $C_6$ roughly scales as $m^{11}$($m$ is the principal quantum number). \textcolor{black}{Details about the influence from original nonresonant dipole-dipole coupled states would be discussed in Appendix B.} Therefore, the reduced {\it vdWs}-type interaction Hamiltonian can be described by
\begin{equation}
     \hat{U}_{c_jc_{j^{\prime}}} =
 \frac{C_6}{R_{c_jc_{j^{\prime}}}^6}|p_jp_{j^{\prime}}\rangle\langle p_jp_{j^{\prime}}|.
 \label{Uccj}
 \end{equation}
 
 From Eqs.(\ref{Uctsp}) and (\ref{Uccj}) it is apparent that both the dipole-dipole and {\it vdWs} interactions between two Rydberg atoms are separation-dependent. To reach a huge asymmetry in the interaction {\it i.e.} $\frac{C_3^{sp}(\theta_{c_jt})}{R_{ct}^3}\gg \frac{C_6}{R_{c_jc_{j^{\prime}}}^6}$, we have to seek for optimal distributions of  all control atoms on the spherical surface, see more details in Sec.\ref{MAX}.

 \section*{Appendix B: Leakage error estimation}
\label{asB}

{\it Leakage error based on two-atom states.} As illustrated in Fig.\ref{diagram}a we consider a resonant dipole-dipole interaction between one control atom and one target atom for the $|p_js_t\rangle\leftrightarrows|s_jp_t\rangle$ transition. In a real implementation these two-atom pair states might still
experience nonresonant dipole-dipole couplings to other undesired Rydberg pair states, resulting in a leakage error to the gate fidelity. Here the nonresonant coupling strength and the F\"{o}rster energy defect are denoted as $B_\kappa$ and $\delta_\kappa$ respectively. Our task is to find out the influence of these nonresonant couplings to the gate fidelity estimated in our protocol.
In principle we should sum over all selection-rule permitted transitions over a wide range of principal quantum numbers and calculate the leakage error. 
Here we have checked all possible transitions from $|p_js_t\rangle$ and $|s_jp_t\rangle$ to other leakage states and find that the influence of a farther state can be almost negligible due to its weaker coupling strength $B_\kappa$ or a larger energy defect $\delta_\kappa$. In the calculation the factor $B_\kappa/\delta_\kappa$ is used to characterize the leakage strength that is proportional to the leakage error. If $B_\kappa/\delta_\kappa\ll1$ the leakage from the resonantly-coupled state can be suppressed \cite{PhysRevA.96.042306}.

In Fig.\ref{diagram}a and Table I(left) we show the possible transitions with the change of principal quantum number up to $\pm2$ from the resonant pair states $|p_js_t\rangle\leftrightarrows|s_jp_t\rangle$.
To estimate the gate error due to the leakage of population from these states, we solve the stochastic Schr\"{o}dinger equation (\ref{she}) with respect to the Hamiltonian
 \begin{eqnarray}
    &\mathcal{H}_{ps}&  \nonumber\\
     &=&\frac{1}{2}\Omega_2(t)|p_j1_t\rangle\langle p_js_t| + \frac{1}{2}\Omega_3(t)|p_js_t\rangle\langle p_j0_t| \nonumber \\
    &+& (B_0|p_js_t\rangle\langle s_jp_t|
    +B_{\kappa}|p_js_t\rangle\langle a_{\kappa j}b_{\kappa t}|+B_{\kappa}|s_jp_t\rangle\langle b_{\kappa j}a_{\kappa t}| \nonumber \\
    &+& H.c.)+\delta_{\kappa}(|b_{\kappa j}a_{\kappa t}\rangle\langle b_{\kappa j}a_{\kappa t}| + |a_{\kappa j}b_{\kappa t}\rangle\langle a_{\kappa j}b_{\kappa t}|),
\end{eqnarray}
where we treat $|p_j1_t\rangle$ as the initial state and apply two pulses $\Omega_2(t)$ and $\Omega_3(t)$. Ideally a pre-excitation of the control atom would block the excitation of the target atom so the population missing from state $|p_j1_t\rangle$ can be regarded as the leakage error. See the last column of Table I(left), the population rotation error shows a clear decrease
 from case 1 to case 4 and it is almost negligible in cases 3 and 4 due to the tiny leakage error in the range of $10^{-7}\sim10^{-5}$. Based on the results we choose two pairs of nonresonant coupled states:
 $|a_{1j}b_{1t}\rangle$ and $|b_{1j}a_{1t}\rangle$($\kappa=1$), $|a_{2j}b_{2t}\rangle$ and $|b_{2j}a_{2t}\rangle$($\kappa=2$) as the dominant leakage states with respect to the resonant transition between $|p_js_t\rangle\leftrightarrows|s_jp_t\rangle$.

\begin{widetext}

\begin{figure}
\centering\includegraphics[width=0.70\textwidth]{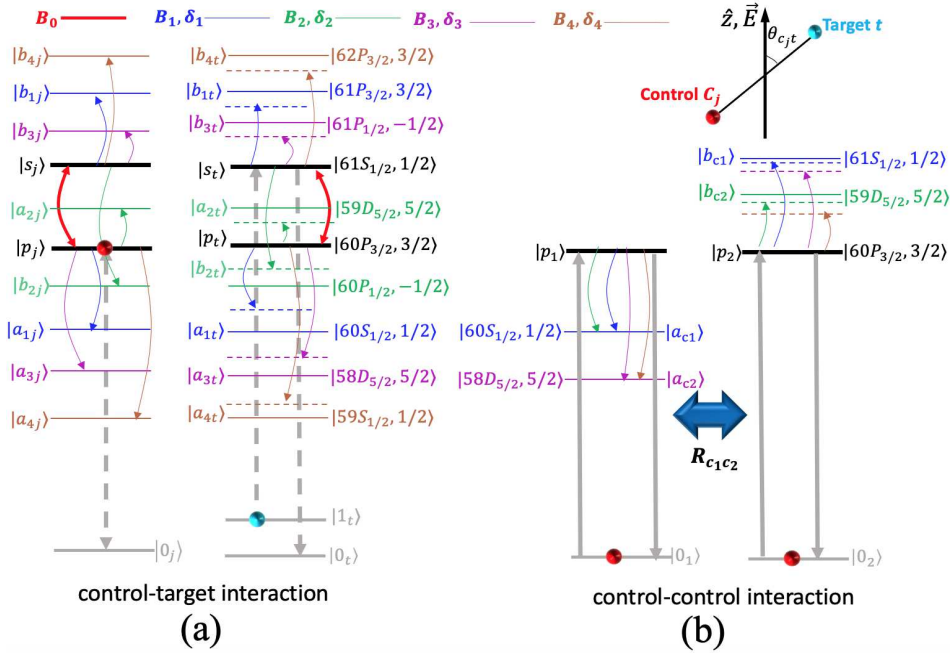}
\caption{(a) For one control and one target atoms, some dominant leakage levels with respect to the resonant dipole-dipole interaction between $|p_js_t\rangle$ and $|s_jp_t\rangle$, where 
$|p_{j,t}\rangle=|60P_{3/2},3/2\rangle$ and $|s_{j,t}\rangle=|61S_{1/2},1/2\rangle$. Other selection-rule permitted pair states are off-resonantly coupled by obeying $|p_js_t\rangle\leftrightarrows|a_{\kappa j}b_{\kappa t}\rangle$(forward) and $|s_jp_t\rangle\leftrightarrows|b_{\kappa j}a_{\kappa t}\rangle$(backward) with strength $B_{\kappa}$ and F\"{o}rster energy defect $\delta_{\kappa}$ and $\kappa\in(1,2,3,4)$.
(b) Some leakage levels related to the detuned dipole-dipole interaction between two control atoms where $|p_1p_2\rangle=|60P_{3/2},3/2;60P_{3/2},3/2\rangle$. $B_{\kappa}$ and $\delta_{\kappa}$ represent the coupling strength and the corresponding energy defect between $|p_1p_2\rangle$ and other undesired nonresonant Rydberg pair states $|Pr_\kappa\rangle=\{|a_{c1}b_{c1}\rangle$, $|a_{c1}b_{c2}\rangle$, $|a_{c2}b_{c1}\rangle$, $|a_{c2}b_{c2}\rangle\}$, where $|a_{c1}\rangle=|60S_{1/2},1/2\rangle$, $|a_{c2}\rangle=|58D_{5/2},5/2\rangle$, $|b_{c1}\rangle=|61S_{1/2},1/2\rangle$, $|b_{c2}\rangle=|59D_{5/2},5/2\rangle$.}
\label{diagram}
\end{figure}
\begin{table*}[!htbp]
\centering
\renewcommand\arraystretch{1.3}
\resizebox{\textwidth}{!}{
\begin{tabular}{c|c c c c c || c c c c c } 
\hline 
$\kappa$ &$|a_{\kappa j}\rangle=|a_{\kappa t}\rangle$&$|b_{\kappa j}\rangle=|b_{\kappa t}\rangle$&$C_{3}^{sp(\kappa)}(0)$&$B_{\kappa}/\delta_{\kappa}$&$ 1-P_{p_{j}1_{t}}$&$ |Pr_{\kappa}\rangle$&$C_{3}^{(\kappa)}$&$\epsilon_{1}$&$\epsilon_{2}$\\
\hline 
1&$|60S_{1/2},\frac{1}{2}\rangle$&$|61P_{3/2},\frac{3}{2}\rangle$&$-9.134$&$8.3\times10^{-2}$&$9.11\times10^{-4}$&$|60S_{1/2},\frac{1}{2};61S_{1/2},\frac{1}{2}\rangle$&$4.301$&$4.27\times10^{-5}$&$3.21\times10^{-2}$\\
\hline 
2&$|59D_{5/2},\frac{5}{2}\rangle$&$|60P_{1/2},-\frac{1}{2}\rangle$&$-9.254$&$9.5\times10^{-3}$&$1.32\times10^{-4}$&$|60S_{1/2},\frac{1}{2};59D_{5/2},\frac{5}{2}\rangle$&$5.919$&$2.69\times10^{-8}$&$8.09\times10^{-6}$\\
\hline 
3&$|58D_{5/2}\frac{5}{2}\rangle$&$|61P_{3/2},-\frac{1}{2}\rangle$&$-3.926$&$3.6\times10^{-3}$&$3.93\times10^{-5}$&$|58D_{5/2},\frac{5}{2};61S_{1/2},\frac{1}{2}\rangle$&$3.203$&$2.15\times10^{-8}$&$8.76\times10^{-6}$\\
\hline 
4&$|59S_{1/2},\frac{1}{2}\rangle$&$|62P_{3/2},\frac{3}{2}\rangle$&$-0.14$&$2.1\times10^{-4}$&$3.57\times10^{-7}$&$|58D_{5/2},\frac{5}{2};59D_{5/2},\frac{5}{2}\rangle$&$4.408$&$1.43\times10^{-8}$&$8.75\times10^{-6}$\\
\hline 
\end{tabular}}
\caption{Leakage error due to the presence of several nonresonant dipole-dipole coupled states nearby. Table(left): 
Leakage states for the
resonant pair states $|p_js_t\rangle$ and $|s_jp_t\rangle$ which are $|p_js_t\rangle\leftrightarrows|a_{\kappa j}b_{\kappa t}\rangle$(forward) and $|s_jp_t\rangle\leftrightarrows|b_{\kappa j}a_{\kappa t}\rangle$(backward) with strength $B_{\kappa}=C_{3}^{sp(\kappa)}/R_{ct}^3$($C_{3}^{sp(\kappa)}$ is in unit of $2\pi\times$GHz$\cdot\mu$m$^3$) and energy defect $\delta_{\kappa}/2\pi=(0.8771,7.8032,8.6142,5.3452)$ GHz. Here $\theta_{c_jt}=0$ and $B_0=C_3^{sp}(0)/R_{ct}^3$ with $C_3^{sp}(0)/2\pi=-8.388$ GHz$\cdot\mu$m$^3$.
$R_{ct}=5$ $\mu$m treats as the radius of sphere and the leakage error $1-P_{p_j1_t}$ is estimated by calculating the population missing from state $|p_j1_t\rangle$. Table(right): Off-resonance pair states $|Pr_{\kappa}\rangle$ with respect to $|p_1p_2\rangle$ {\it i.e.} $|p_1p_2\rangle\leftrightarrows|Pr_{\kappa}\rangle$ enabled by a nonresonant dipole-dipole interaction $B_\kappa=C_3^{(\kappa)}/R_{c_1c_2}^3$($C_{3}^{(\kappa)}$ is in unit of $2\pi\times$GHz$\cdot\mu$m$^3$). 
Here the energy defect is $\delta_{\kappa}/2\pi=(0.2784,7.0614,7.4587,14.8012)$ GHz and $\epsilon_{1,2}$ stand for the leakage error from state $|0_10_2\rangle$ under different interatomic distances $R_{c_1c_2} = (2R_{ct},R_{ct})$ $\mu$m. All parameters are taken from the ARC open-source library \cite{SIBALIC2017319}.
}
\label{tableps}
\end{table*}

\end{widetext}

 In addition we study the leakage error due to the nonresonant dipole-dipole couplings with respect to the identical Rydberg pair state $|p_1p_2\rangle$ for two control atoms. Here we apply four pairs of dipole-allowed off-resonantly coupled states $|Pr_{\kappa}\rangle$ with strength $B_{\kappa}=C_3^{(\kappa)}/R_{c_1c_2}^3$ and F\"{o}rster energy defect $\delta_{\kappa}$, see Table I(right). In principle we have to sum up all possible nonresonant states at the same time which leads to a second-order level shift of $|p_1p_2\rangle$ represented by a coefficient $C_6\approx\sum_\kappa(C_3^{(\kappa)})^2/\delta_\kappa$.
 
 In the calculation we again calculate the dynamics of a two-atom state $|0_10_2\rangle$ by following the Hamiltonian of
 \begin{equation}
   \mathcal{H}_{pp} = \sum_j^2\mathcal{H}_j+  (B_\kappa|p_1p_2\rangle\langle Pr_\kappa|+H.c.)+\delta_\kappa|Pr_\kappa\rangle\langle Pr_\kappa| 
 \end{equation}
 where $\mathcal{H}_j=\frac{1}{2}(\Omega_{1}(t)|p_j\rangle\langle 0_j|+
\Omega_{5}(t)|0_j\rangle\langle p_j|)$.
Starting from the initial state $|0_10_2\rangle$ we numerically estimate the population missing from $|0_10_2\rangle$($\epsilon_{1(2)}=1-P_{0_10_2}$) by applying two separated $\pi$ pulses $\Omega_1(t)$ and $\Omega_5(t)$. Note that the nonresonant coupling strength $B_{\kappa}$ is inversely proportional to the interatomic distance $ R_{c_1c_2}$ so we use different values
$R_{c_1c_2}=(2R_{ct},R_{ct})$ $\mu$m. See the last two columns in Table I, for a larger distance the leakage error $\epsilon_1$ is always below $10^{-4}$ in all cases which means these off-resonantly coupled levels play a negligible effect and it is reasonable to assume a pure {\it vdWs} shift represented by $C_6/R_{c_1c_2}^6$ for the $|p_1p_2\rangle$ state. However if the two-atom distance is too small {\it e.g.} $R_{c_1c_2}=R_{ct}$, it is inappropriate to assume a {\it vdWs} interaction because the pair state $|Pr_1\rangle$ can cause a big leakage error $\sim0.0321$ which is even larger than the intrinsic decay error. 
In all cases we find the effect of other nonresonant states $|Pr_{2\sim4}\rangle$ can be ignored as compared with $|Pr_1\rangle$. 
Based on the analysis above we treat $|Pr_1\rangle$ as the dominant leakage state with respect to $|p_1p_2\rangle$ and will include it in evaluating the leakage error of a realistic multiqubit quantum gate.

{\it Leakage error based on a multiqubit quantum gate.} To correctly show the results reasonable in the main text, we now discuss the real implementation of a multiqubit quantum gate by taking account of off-resonant dipole-dipole couplings from dominant leakage states. The system we simulate consists of $n$ control atoms and one target atom described by the Hamiltonian of
\begin{equation}
    \hat{\mathcal{H}} = \hat{\mathcal{H}}_0+ \hat{\mathcal{H}}_I.
    \label{HH}
\end{equation}
Here the first term $\hat{\mathcal{H}}_0$ as shown in Eq.(\ref{single}) describes the resonant atom-light couplings of control and target atoms. The second term $\hat{\mathcal{H}}_I$[see Eq.(\ref{inter})] describes the Rydberg states and their interactions, where the resonant dipole-dipole interaction for control-target atom pairs is(here $\kappa=1,2$)
\begin{eqnarray}
    \hat{U}_{c_jt} &=& (B_0|p_js_t\rangle\langle s_jp_t|+B_{\kappa}|p_js_t\rangle\langle a_{\kappa j}b_{\kappa t}| \nonumber\\
    &+&B_{\kappa}|s_jp_t\rangle\langle b_{\kappa j}a_{\kappa t}| + H.c.)
    +\delta_{\kappa}(|b_{\kappa j}a_{\kappa t}\rangle\langle b_{\kappa j}a_{\kappa t}| \nonumber\\
    &+& |a_{\kappa j}b_{\kappa t}\rangle\langle a_{\kappa j}b_{\kappa t}|), 
\end{eqnarray}
and the nonresonant dipole-dipole interaction between two identical control atoms takes form of(here $\kappa=1$)
\begin{equation}
    \hat{U}_{c_jc_{j^{\prime}}} = (B_\kappa|p_1p_2\rangle\langle Pr_\kappa|+H.c.)+\delta_\kappa|Pr_\kappa\rangle\langle Pr_\kappa|, 
\end{equation}
with the strength $B_{\kappa}$ and the F\"{o}rster energy defect $\delta_\kappa$ estimated in Table I. Note that the coefficient $C_3^{sp(\kappa)}$ depends on a varying polarizing angle $\theta_{c_jt}$ while in Table I we set $\theta_{c_jt}=0$.

In the numerical simulation we first consider $n=2$ which is a three-qubit Toffoli gate showing a linear structure in space. We calculate the average gate fidelity $\bar{\mathcal{F}}_2$[see Eq.(\ref{fide})] over eight input states $|\Psi_{in}\rangle=\{|000\rangle,|001\rangle,|010\rangle,|011\rangle,|100\rangle,|101\rangle,|110\rangle,|111\rangle\}$ by evolving the stochastic Schr\"{o}dinger equation.
The output state $|\bar{\Psi}_{out}\rangle$ is obtained at $t=t_{det}$ after 500 stochastic evolutions. Via taking account of the dominant leakage states we finally find the average gate fidelity is $\bar{\mathcal{F}}_2=0.99793741$. As compared with $\bar{\mathcal{F}}_2=0.99832354$ obtained in the case of no nonresonant states, this value is slightly decreased by $3.9\times 10^{-4}$. The small leakage error caused by nonresonant couplings, 
means that the asymmetric blockade interactions can be well preserved in the implementation of a $(2+1)$-qubit gate. 

More remarkably, we also calculate a realistic spheroidal quantum gate with four control atoms and one target atom since different atomic distances and polarizing angles must be affected. After a computer-demanding calculation we finally obtain $\bar{\mathcal{F}}_{4}=0.99360269$ which leads to a leakage error of $7.1\times10^{-4}$ with respect to the case that ignores the nonresonant couplings. In Fig.\ref{fed} we have shown that the average gate fidelity for the case of $n=4$ is 0.99431526. This slight growth of the leakage error with the number of control atoms mainly comes from a slightly weaker asymmetry in the interaction. Because as $n$ increases, the nonresonant coupling strength which depends on the interatomic distance $R_{c_jc_j^\prime}$, becomes stronger arising a bigger leakage. However our simulations confirm that, by considering off-resonantly coupled Rydberg states the leakage error for a realistic 3D multiqubit quantum gate can still be kept at a negligible level $\sim10^{-4}$. Therefore the 3D multiquibit gate presented in the main text can be implemented in a realistic environment which is enabled by a large symmetry in the interaction.

\section*{Appendix C: Evolutionary algorithm}

\begin{table*}[!htbp]
\centering
\begin{tabular}{c | c c c c c c} 
\hline 
 Case&$c_1$&$c_2$&$c_3$&$c_4$&$c_5$&$c_6$\\
\hline  
(i) &$(\pi/2,\pi/6)$&$(\pi/2,\pi/2)$&$(\pi/2,5\pi/6)$&$(\pi/2,7\pi/6)$&$(\pi/2,3\pi/2)$&$(\pi/2,11\pi/6)$\\
  &$(1.5731,3.1421)$&$(1.5747,4.7111)$&$(1.5732,0)$&$(1.5705,1.5694)$&$(0.0027,0.0007)$&$(3.1412,0.0012)$\\
\hline 
(ii) &$(0,0)$&$(\pi/5,2\pi/5)$&$(2\pi/5,4\pi/5)$&$(3\pi/5,6\pi/5)$&$(4\pi/5,8\pi/5)$&$(\pi,2\pi)$\\
 &$(1.5743,3.1433)$&$(1.5744,4.7136)$&$(1.5722,0)$&$(1.5724,1.5753)$&$(0.0003,0.0008)$&$(3.1408,0.0056)$\\
\hline
(iii) &$(\pi/6,\pi/5)$&$(\pi/3,2\pi/5)$&$(\pi/2,3\pi/5)$&$(2\pi/3,4\pi/5)$&$(5\pi/6,\pi)$&$(\pi,6\pi/5)$\\
 &$(1.5701,3.1403)$&$(1.5697,4.7125)$&$(1.5711,0)$&$(1.5707,1.5731)$&$(0.0010,0.0011)$&$(3.1419,0.0023)$\\
\hline 
(iv) &$(\pi/7,\pi/5)$&$(2\pi/7,\pi/6)$&$(3\pi/7,\pi/7)$&$(4\pi/7,\pi/8)$&$(5\pi/7,\pi/9)$&$(6\pi/7,\pi/10)$\\
 &$(1.5705,3.1419)$&$(1.5711,4.7119)$&$(1.5695,0)$&$(1.5703,1.5718)$&$(0.0002,0.0005)$&$(3.1416,0.0003)$\\
\hline 
(v) &$(\pi/8,\pi)$&$(\pi/4,\pi/2)$&$(3\pi/8,\pi/3)$&$(\pi/2,\pi/4)$&$(5\pi/8,\pi/5)$&$(3\pi/4,\pi/6)$\\
 &$(1.5713,3.1408)$&$(1.5716,4.7119)$&$(1.5701,0)$&$(1.5695,1.5705)$&$(0.0013,0.0007)$&$(3.1415,0.0001)$\\
\hline 
Average results &$(1.5708,3.1416)$&$(1.5707,4.7124)$&$(1.5709,0)$&$(1.5708,1.5707)$&$(0.0005,0.0006)$&$(3.1416,0.0007)$\\
\hline 

\end{tabular}
\caption{ (i-v) Five sets of parameters for random initialization $(\theta_j^{(0)},\phi_j^{(0)})$ of atomic positions(first row) and for optimized positions($(\theta_j,\phi_j)$, second row) under one optimization run. Average results indicate the datum averaging over $10^4$ optimizations. }
\label{table1}
\end{table*}

In this section we demonstrate how to get optimal geometries of control atoms on the surface and determine the value $n_{max}$ via evolutionary algorithm. 
Taking C$_6$NOT gate as an example, first we randomly arrange the initial positions of control atoms $c_j$ denoted as $\bold{r}_j^{(0)}=(R_{ct},\theta_j^{(0)},\phi_j^{(0)})$ where the superscript ``0" means the initial step $p=0$. For a finite spherical radius $R_{ct}$, one has $\bold{r}_j^{(0)}=(\theta_j^{(0)},\phi_j^{(0)})$ and $j\in[1,6]$. Then we compare all asymmetry factors $\chi_j^{(0)}=U_{c_jt}/U_{c_jc_{j^{\prime}}}$ with respect to atom $c_j$ where the total number is C$_6^2=15$, in order to find a minimal value
\begin{equation}
    \chi_0= \min(\chi_j^{(0)}).
\end{equation}
 
 Next we add a small perturbation to the position of control atoms, leading to
 \begin{equation}
   \bold{r}_j^{(1)}=(\theta_j^{(0)}+\delta\theta_j,\phi_j^{(0)}+\delta\phi_j),  
 \end{equation}
in which $\delta\theta_j$ or $\delta\phi_j$ is obtained randomly from the range of $[-0.1\theta_j^{(0)},0.1\theta_j^{(0)}]$ or $[-0.1\phi_j^{(0)},0.1\phi_j^{(0)}]$. With the new position $ \bold{r}_j^{(1)}$ we again compare all $\chi_j^{(1)}$ values and find out a minimal value which is denoted as $\min(\chi_j^{(1)})$. To maximize asymmetric blockade, $\chi_1$ is defined  as
\begin{equation}
    \chi_1= \max(\min(\chi_j^{(0)}),\min(\chi_j^{(1)})),
\end{equation}
at step $p=1$ and the corresponding position $\bold{r}_j^{(0)}$ or $\bold{r}_j^{(1)}$ will be ready for the next-step($p=2$) optimization. This single-optimization process must be repeated with sufficient iterations(typically $p>10^5$) until the condition $|\chi_p-\chi_{p-1}|< 10^{-5}$ is met where the universal maximal asymmetry factor is given by
\begin{equation}
    \chi_p= \max(\min(\chi_j^{(0)}),\min(\chi_j^{(1)}),...\min(\chi_j^{(p)})).
\end{equation}

In order to avoid a local optimal solution we perform $10^4$ optimization runs via evolutionary algorithm and achieve the optimal distribution of control atoms $\bold{r}_j=(\theta_j,\phi_j)$. As shown in Fig.\ref{mod}(a) the position of $c_2$ is obtained after $10^4$ optimizations which are very condensed in space. To quantitatively verify this effect, in Table \ref{table1}(i-v) we exemplify five sets of parameters to show the robustness of our optimization algorithm. Given the initial positions $(\theta_j^{(0)},\phi_j^{(0)})$ (first row of (i-v)), the results from single optimization(second row of (i-v)) are very close to the average results after taking $10^4$ optimizations, which confirms the accuracy of evolutionary algorithm. For ensuring a huge asymmetry we also set a limitation $\chi_p>100$ in the optimization which arises  a maximal $n_{max}$ permitted if $m$ and $R_{ct}$ are determined. E.g. in the case of $m=60$ and $R_{ct}=5$ $\mu$m, $n=8$ leads to $\chi_p=102.986$, while $n=9$ leads to $\chi_p=64.310$ which breaks the limitation, so $n_{max}=8$ is obtained.


\section*{Appendix D: Optimal geometry}

\begin{figure*}
\centering
\includegraphics[width=0.75\textwidth]{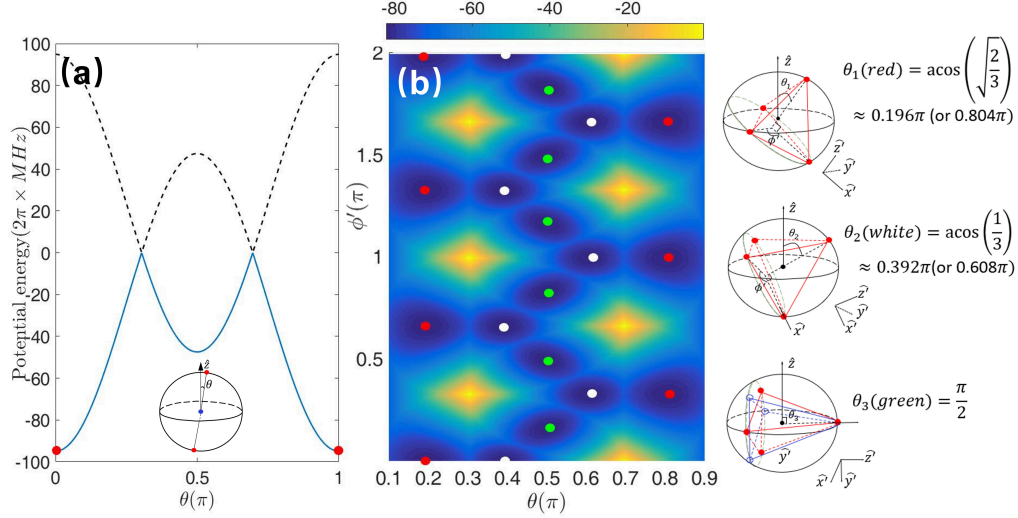}
\caption{Collective potential eigenenergies $E_{\pm}$ and $E_{2,-}$ corresponding to (a) two and (b) four control atoms, {\it vs} the polarizing angle $\theta$ and the polarizing and azimuthal angles ($\theta,\phi^{\prime}$), respectively. Subscript $j$ is omitted for brevity. (b) In the case of four control atoms we introduce another angle $\phi^{\prime}$ with respect to the rotational coordinate of $(\hat{x}^{\prime},\hat{y}^{\prime},\hat{z}^{\prime})$, in order to determine the spatial position of atoms. On the right side of (b) we show three possible distributions of control atoms in a 3D sphere corresponding to red dots, white dots and green dots,
where the absolute value of potential energy $E_{2,-}$ attains its maximum. }
\label{twop}
\end{figure*}

{\it Case of two control atoms.} To provide a physical understanding for these optimal configurations we study the potentials of system.
The basic ingredient in optimization is to maximize the asymmetry of interaction, arising $U_{c_jt}\gg U_{c_jc_j^\prime}$. To achieve this, we explore the essential feature of system by concentrating on the Rydberg states $|p_{j,t}\rangle$, $|s_{j,t}\rangle$ which connect with the interaction Hamiltonians Eqs.(\ref{uct}-\ref{ucc}). Here the index $j=1,...,n$ represents the number of control atoms, and $t$ labels the target atom. We analyze the interaction potential in the subspace only involving Rydberg states spanned by collective states $|\pi_i\rangle$($i=1,...,n$), which means $i$ control atoms are in $|p\rangle$ and others are in $|s\rangle$ \cite{PhysRevA.89.053426}.
For $n=2$, states $|\pi_i\rangle$ can be written as
\begin{eqnarray}
 |\pi_1\rangle &=& \frac{1}{\sqrt{2}}(|s_1p_2p_t\rangle + |p_1s_2p_t\rangle), \nonumber \\
 |\pi_2\rangle &=& |p_1p_2s_t\rangle.
 \label{pistat}
\end{eqnarray}
Due to the exchange property of resonant dipole-dipole interaction $|p_js_t\rangle\leftrightarrows|s_jp_t\rangle$ this complete set of basis is equivalent to the form of  $\{|\pi_0\rangle,|\pi_1\rangle\}=\{|s_1s_1p_t\rangle,\frac{1}{\sqrt{2}}(|s_1p_2s_t\rangle + |p_1s_2s_t\rangle)\}$. In the basis of $|\pi_1\rangle$ and $|\pi_2\rangle$ as in Eq.(\ref{pistat}), the effective interaction Hamiltonian $\mathcal{H}_{n,int}$ is given by ($n=2$)
\begin{equation}       
\mathcal{H}_{2,int}=\left(                 
\begin{array}{ccc}   
0 & \frac{1}{\sqrt{2}}D_2\\  
\frac{1}{\sqrt{2}}D_2 & B_2\\ 
\end{array}
\right),    
\label{dia}
\end{equation}
where $D_n = \sum_{j=1}^n{U_{c_jt}}$ and $B_n = \sum_{j>j^{\prime}}{U_{c_jc_{j^{\prime}}}}$ and $j,j^{\prime}\in[1,..,n] $. 
Analytical diagonalization of the Hamiltonian matrix yields the eigenenergies, explicitly as
\begin{equation}
    E_{\pm} = \frac{B_2\pm\sqrt{B_2^2+2D_2^2}}{2}.
\end{equation}

It is apparent that, in Eq.(\ref{dia}) the off-diagonal resonant dipole-dipole interaction $U_{c_jt}(\theta_{c_jt})\propto (1-3\cos^2(\theta_{c_jt}))$ is responsible for the population transfer between different $|\pi_i\rangle$ states, which leads to anisotropic spatial interactions. The diagonal term $U_{c_1c_2}\propto|\bold{r}_{c_1}-\bold{r}_{c_2}|^{-6}$ also depends on the relative distance between two control atoms $c_1$ and $c_2$. We restrict all control atoms on the spherical surface so $U_{c_1c_2}$ is minimized when two atoms are separated by a maximal distance which is $2R_{ct}$(diameter), leading to $\theta=\theta_{c_1t}=\pi-\theta_{c_2t}$. In this case $D_2$ and $B_2$ take explicit expressions as
\begin{equation}
    D_2=\frac{2C_3(1-3\cos^2\theta)}{R_{ct}^3},B_2=\frac{C_6}{(2R_{ct})^6},
\end{equation}
and the eigenenergies of two collective states $|\pi_{1,2}\rangle$ can be independently controlled by a single polarizing angle $\theta$.

As shown in Fig. \ref{twop}(a), we show the eigenenergies $E_{\pm}$ as a function of the polarizing angle $\theta$. For 
$\theta=0$ or $\pi$ the minimum of potential occurs which means the attractive dipole-dipole interaction $D_2$ attains its maximal(absolute) magnitude. In other word when the two control atoms are placed at $(\theta,\phi)=(0,0)$ and $(\pi,0)$ the asymmetric interaction could be maximized (note that $B_2$ is a constant).
Our result based on numerical optimization also gives to a same structure as the theoretical prediction, see Fig.\ref{mod}(n=2).
To this end we confirm the accuracy of evolutionary algorithm which helps to obtain a maximal asymmetric interaction by positioning atoms appropriately on the spherical surface.

{\it Case of four control atoms.} Situation becomes quite complex when more control atoms are included, because the magnitude of potential energy is influenced by multiple adjustable parameters. As examples we analyze the case of $n=4$. With the inclusion of more Rydberg states the collective $|\pi_i\rangle$ states can be rewritten as 
\begin{eqnarray}
 |\pi_1\rangle &=& \frac{1}{2}(|p_1s_2s_3s_4p_t\rangle+|s_1p_2s_3s_4p_t\rangle+|s_1s_2p_3s_4p_t\rangle \nonumber  \\
 &+&|s_1s_2s_3p_4p_t\rangle), \nonumber \\
 |\pi_2\rangle &=& \frac{1}{\sqrt{6}}(|p_1p_2s_3s_4s_t\rangle+|p_1s_2p_3s_4s_t\rangle+|p_1s_2s_3p_4s_t\rangle
  \nonumber \\ 
 &+&|s_1p_2p_3s_4s_t\rangle+|s_1p_2s_3p_4s_t\rangle+|s_1s_2p_3p_4s_t\rangle), \nonumber\\
 |\pi_3\rangle &=& \frac{1}{2}(|p_1p_2p_3s_4p_t\rangle+|p_1p_2s_3p_4p_t\rangle+|p_1s_2p_3p_4p_t\rangle \nonumber\\
 &+&|s_1p_2p_3p_4p_t\rangle), \nonumber\\
 |\pi_4\rangle &=& |p_1p_2p_3p_4s_t\rangle, 
\end{eqnarray}
arising the effective interaction Hamiltonian $\mathcal{H}_{4,int}$ given by
\begin{equation}       
\mathcal{H}_{4,int}=\left(                 
\begin{array}{cccc}   
0 & \frac{\sqrt{3}}{2\sqrt{2}}D_4 & 0 & 0\\  
\frac{\sqrt{3}}{2\sqrt{2}}D_4 & \frac{1}{6}B_4 &  0 & 0\\ 
0& 0 & \frac{1}{2}B_4 & \frac{1}{2}D_4\\ 
0 & 0 & \frac{1}{2}D_4 & B_4\\  
\end{array}
\right),   
\label{fh}
\end{equation}
with its eigenenergies explicitly as
\begin{eqnarray}
    E_{1,\pm} = \frac{3B_4\pm\sqrt{B_4^2+4D_4^2}}{4}, \\
    E_{2,\pm} = \frac{B_4\pm\sqrt{B_4^2+216D_4^2}}{12}.
\end{eqnarray}

However due to multiple adjustable parameters from four control atoms(there are overall eight degrees of freedom $(\theta_{1\sim4},\phi_{1\sim4})$ in the bare spherical coordinate) it is difficult to find out its maximal magnitude which corresponds to a maximal asymmetric interaction. By considering the symmetry of potentials we guess a regular-tetrahedron structure. As displayed in Fig. \ref{twop}(b)(right inset) an optimal regular-tetrahedron configuration is freely rotated in which the polarizing angle $\theta$(denoted by $\theta_1$, $\theta_2$, $\theta_3$) with respect to one control atom(vertex) varies from 0 to $\pi/2$. Note that the range of $\theta\in(\pi/2,\pi)$ is equivalent due to the symmetry so we focus on $\theta\in(0,\pi/2]$.
Besides, $\phi^{\prime}$ is an azimuthal angle in the bottom side of tetrahedron which is opposite to the vertex. With tunable $\theta$ and $\phi^{\prime}$, we show the potential energy $E_{2,-}$ in Fig. \ref{twop}(b) where its maximal(absolute) values are denoted by dots. Periodical patterns along $\phi^{\prime}$ axis are explicitly observed due to the isotropy of three atoms in the bottom side. However, only at $(\theta_1,\theta_2,\theta_3)=(\arccos (\sqrt{\frac{2}{3}}),\arccos (1/3),\pi/2)$, the absolute value of $E_{2,-}$ which is $\propto |D_4|$ can reach its global maximum, corresponding to the optimal configurations as displayed on the right side of Fig. \ref{twop}(b). Note that for $\theta_3=\pi/2$ there exists two types of bottom-side atomic distributions due to the periodicity.

To obtain a maximal asymmetric blockade among the cases of $\theta_1$, $\theta_2$ and $\theta_3$, we analyze  the ratio $\chi_j$ of each control atom and find that its minimum is $\min(\chi_j)\approx828.44$ when $\theta=\theta_1$(case 1). And this value lowers to 552.30 if $\theta=\theta_2$(case 2). Case 3 with $\theta=\theta_3$ has an equivalent structure as the former case 2. Relevant parameters estimated in calculation are $C_3/2\pi=4.194$ GHz$\cdot\mu$m$^3$, $C_6/2\pi=-12$ GHz$\cdot\mu$m$^6$ and $R_{ct}=5.0$ $\mu$m. These results explicitly suggest that case 1 has a best asymmetry, which agrees with the optimal structure obtained by evolutionary algorithm, see Fig.\ref{mod}(n=4).

\section*{Appendix E: Gate Implementation}

 In the section we demonstrate how to realize a C$_n$NOT gate via asymmetric blockade.
 Consider $^{87}Rb$ atoms trapped in a 3D spheroidal array with $R_{ct}=5.0$ $\mu$m. Qubit-state preparation depends on two hyperfine ground states  $|0_{j,t}\rangle=|5S_{1/2},F=1,m_F=0\rangle$ and $|1_{j,t}\rangle=|5S_{1/2},F=2,m_F=0\rangle$ via an optically pumping method \cite{Kim:18}. In practice the control atoms are globally driven via an one-step UV excitation from $|0_j\rangle$ to $|p_j\rangle=|60P_{3/2}\rangle$
 with wavelength 297nm; and the target atom will face a two-photon transition from $|0_t\rangle$ to $|s_t\rangle=|61S_{1/2}\rangle$
 with wavelengths 795nm and 474nm, decided by the selection rules. For the target atom the intermediate state {\it e.g.} $|5P_{1/2}\rangle$ has been safely discarded due to dispersive interactions, and state $|1_t\rangle$ is also coupled to $|s_t\rangle$ by same mechanism.

 A straightforward realization of the multiqubit C$_n$NOT gates should obey
 \begin{eqnarray}
 |00...0...\beta\rangle &=&  e^{i\varphi}|00...0...\beta\rangle \nonumber \\
\text{and } |\underbrace{11...1...}_{n}\beta\rangle &=& |\underbrace{11...1...}_{n}\bar{\beta}\rangle
 \end{eqnarray}
with $\beta\in[0,1]$ and $\bar{\beta}\equiv1-\beta$. We have omitted the subscripts {\it t, j}. The relative phase $\varphi$ is tunable by external fields and here $\varphi=0$. To describe the gate implementation, we note that when any control atom is initialized in state $|0_j\rangle$ the pre-excitation of the control qubit(s) can induce a strong control-target exchange interaction which blocks the subsequent excitation of the target atom. Only if all control atoms are in idle state $|1_j\rangle$ that are uncoupled with the laser fields, a complete state conversion mediated by $|s_t\rangle$ takes place
between $|0_t\rangle$ and $|1_t\rangle$ of the target atom. In the scheme, a weak control-control interaction can facilitate an individual manipulation for the target atom by each control qubit, avoiding the control-control blockade.

As usual we apply a piecewise pulse sequence $\Omega_{1\sim5}$ comprising three fundamental steps \cite{shi2022quantum}. (1) The incidence $\pi$ pulse with Rabi frequency $\Omega_1$ is globally applied to all control atoms, which allows a Rydberg excitation of $|0_j\rangle\to|p_j\rangle$. (2) A pair of $\pi$ pulses which include 
$\Omega_{2}$ and $\Omega_{3}$(or $\Omega_{3}$ and $\Omega_{4}$), can generate a state swap for the target atom between $|0_t\rangle$ and $|1_t\rangle$ if all control atoms are idle in $|1_j\rangle$, obeying the routes of
\begin{eqnarray}
 |1_t\rangle\xrightarrow{\Omega_2}|s_t\rangle\xrightarrow{\Omega_3}|0_t\rangle,  \nonumber\\
 \text{or }|0_t\rangle\xrightarrow{\Omega_3}|s_t\rangle\xrightarrow{\Omega_4}|1_t\rangle, 
 \label{swap}
\end{eqnarray}
 which depends on its initial status $|1_t\rangle$(or $|0_t\rangle$). (3) A $(-\pi)$-pulse $\Omega_5$ can finally return the Rydberg state $|p_j\rangle$ to $|0_j\rangle$ via a de-excitation process. In the main text we have assumed the Rabi frequencies with magnitudes $\Omega_c = |\Omega_{1,5}|$ and $\Omega_t = |\Omega_{2,3,4}|$ throughout the paper.

\section*{Appendix F: Other technical errors}

\begin{figure}
\centering
\includegraphics[width=0.48\textwidth]{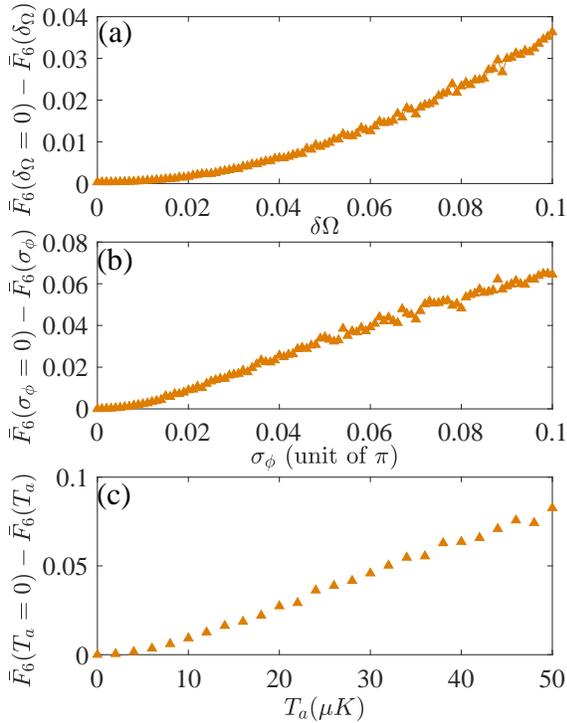}
\caption{ Technical errors of a C$_6$NOT gate caused by the imperfection of experimental conditions. (a-b) Deviations in the laser amplitudes and phases. (c) Motional dephasing of atoms under different atomic temperatures. Every point represents an average over 500 measurements.}
\label{Figterror}
\end{figure}

In this section we first address the errors from imperfect experimental technique leading to unwanted laser amplitude and phase fluctuations. The infidelity of the C$_6$NOT gate {\it vs} fluctuations of laser field amplitude is plotted in Fig.\ref{Figterror}(a) in which the fluctuation $\delta\Omega_{c(t)}$ with respect to $\Omega_{c(t)}$ is estimated by a random number. {\it i.e.} The relative variation of laser Rabi frequency $\delta\Omega_{c(t)}/\Omega_{c(t)}$ can be randomly obtained from a range of $[-\delta_{\Omega},\delta_{\Omega}]$. By increasing the value $\delta_{\Omega}$, from Fig.\ref{Figterror}(a) we know that the gate infidelity has an apparent growth. For $\delta_\Omega = 0.1$ the infidelity of a 3D C$_6$NOT gate is about $\sim0.035$ after an average of 500 measurements. Similarly we also study the robustness of gate against the variation of laser phases. 
We assume that the fluctuated laser phase $\phi_{c(t)}$ obeys a Gaussian distribution around the desired value $\phi_{c(t),0}=0$ with the standard deviation $\sigma_{\phi}/\pi\in[0,0.1]$. During each measurement a random laser phase $\phi_{c(t)}$ is adopted which leads to $\Omega_{c(t)}\to\Omega_{c(t)}e^{i\phi_{c(t)}}$. From Fig.\ref{Figterror}(b) we learn that the imperfection of gate fidelity also increases with the strength of phase fluctuation $\sigma_{\phi}$. In fact once the laser amplitude or phase is fluctuated it would arise imperfect excitation or deexcitation during the process of state swap owing to the invalidity of $\pi$-pulses [see Eq.(\ref{swap})] which will cause the gate inefficient.

Another inevitable technical error resource is the
motional dephasing effect \cite{PhysRevApplied.13.024008,PhysRevA.97.053803}. Due to the finite temperature, the thermal motion of control and target atoms will induce an inevitable Doppler dephasing to the excitation of Rydberg states which can be estimated by a phase change to the laser Rabi frequencies
\begin{equation}
    \Omega_t \to \Omega_te^{i\Delta_t t}, \Omega_c \to \Omega_ce^{i\Delta_c t},
\end{equation}
where the detuning $\Delta_{t(c)}$ seen by the atoms satisfies a Gaussian distribution with its mean value $\bar{\Delta}_{t(c)}=0$ and the standard deviation $\sigma_{\Delta_{t(c)}}$. Typically $\sigma_{\Delta_{t(c)}} = \bold{k}_{eff}\bold{v}$ where $\bold{k}_{eff}=\sum_j\bold{k}_j$ is the overall wavevector and $\bold{v}=v_{rms}=\sqrt{k_BT_a/M}$ is the atomic root-mean-square velocity with $k_B$, $T_a$
and $M$ being the Boltzmann constant, atomic temperature, and atomic mass. To minimize the Doppler effect, for target atom which undergoes a two-photon transition with wavelengths 795nm and 474nm the effective wavevector is $k_{t,eff}=(k_{474}-k_{495})=5\times 10^6$ $m^{-1}$ by considering two counterpropagating lasers, leading to $\sigma_{\Delta_{t}}=k_{t,eff}v_{rms}$ \cite{PhysRevLett.113.053602}. As for control atoms undergoing an one-step UV excitation $k_{t,eff}=k_{295}=2\times 10^7$ $m^{-1}$ arising $\sigma_{\Delta_{c}}=k_{c,eff}v_{rms}$.
With these settings, in Fig. \ref{Figterror}(c) we numerically calculate the imperfection of gate performance with respect to the temperature. For a given $T_a$ we adopt a random detuning $\Delta_{t(c)}$ from the Gaussian function to simulate the phase error on the population evolution. By averaging over sufficient measurements we find the gate infidelity attains 0.08 at $T_a=50$ $\mu$K where the phase variations caused by atomic thermal motion are $\sigma_{\Delta_t} = 0.346$ MHz, $\sigma_{\Delta_c} = 1.382$ MHz respectively.

\bigskip

\bigskip

\bibliography{references}


\end{document}